\documentclass[12pt]{article}
  \usepackage{amsfonts}
\usepackage[centertags]{amsmath}
\usepackage{amssymb}
\usepackage{graphicx}
\newtheorem{theorem}{Theorem}[section]

\newtheorem{definition}[theorem]{Definition}

\usepackage[verbose,colorlinks=true,naturalnames=true,linkcolor=blue,]{hyperref}

\newcounter{rown}

\begin{document}

\title{New twisted quantum deformations of \\$D=4$ super-Poincar\'{e} algebra\footnote{The talk
given by third author at the International Workshop ''Supersymmetries and Quantum
Symmetries'' (SQS'07) (Dubna, July 30 - August 4, 2007) }}

\author{A. Borowiec$^{1,3}$, J. Lukierski$^{1}$ and V.N. Tolstoy$^{1,2}$
\\ \\
$^{1}$Institute for Theoretical Physics,
\\University of Wroc{\l}aw, pl. Maxa Borna 9,
\\50--205 Wroc{\l}aw, Poland\\
\\$^{2}$Institute of Nuclear Physics,
\\Moscow State University, 119 992 Moscow, Russia\\
\\$^{3}$Bogoliubov Laboratory of Theoretical Physics,
\\Joint Institute for Nuclear Research, Dubna,
\\ Moscow region 141980, Russia}


\date{}
\maketitle
\begin{abstract}
We show how some classical $r$-matrices for the $D=4$ Poincar\'{e} algebra can be
supersymmetrized by an addition of part depending on odd supercharges. These
$r$-mat\-rices for $D=4$ super-Poincar\'{e} algebra can be presented as a sum of the
so-called subordinated $r$-matrices of super-Abelian and super-Jordanian type.
Corresponding twists describing quantum deformations are obtained in an explicit form.
These twists are the super-extensions of twists obtained in the paper \cite{T1}
(\url{arXiv:math/0712.3962v1}). 

\end{abstract}

\section{Introduction}

In the theory of fundamental interactions and field-theoretical models the following two
basic ideas were considered:\\
(i) Since more than thirty years one studies the {\it supersymmetric extensions} of
standard symmetries, dynamical models and classical geometries (see e.g.
\cite{WZ} -- \cite{Buch}).\\
(ii) In last twenty years the idea of introducing {\it quantum symmetries} \cite{D1} -
\cite{FRT} and quantum deformations of dynamical models (see e.g. \cite{KLM,FKN}) is
gaining popularity as a way to describe the effects related with quantization of gravity
\cite{DFR,Gar} and the description of D-branes defined on noncommutative space-time
manifolds \cite{Chu,SW}.

In this paper we shall investigate the superposition of these two concepts in the area of
Hopf-algebraic description of relativistic symmetries. The first example of
supersymmetric extension of quantum symmetries were related with the algebra
$\mathfrak{osp}(1|2)$ \cite{Kulish,KR}; further the supersymmetric extension of
Drinfeld-Jimbo deformation of arbitrary Lie algebra was considered \cite{CK,KT1}.  The
supersymmetric extension of $D=4$ quantum relativistic symmetries was firstly obtained as
the supersymmetrization of $\kappa$-deformed Poincare symmetries \cite{LNRT} --
\cite{KLMS2}; only quite recently new results were obtained as an extension of canonical
Abelian twist for Poincare symmetries \cite{CKNT,Wess} to the case of super-Poincare
symmetries (see e.g. \cite{KS,Zup,BLS}). It should be stressed that recently there was
introduced also general classification of twisted quantum deformations of semisimple Lie
superalgebras \cite{T3,T4}. Subsequently these general considerations were applied to the
quantum deformation of $\mathfrak{osp}(1|2)$ \cite{BLT} which can be interpreted as $D=1$
superconformal symmetry and recently to quantum deformation of $\mathfrak{osp}(1|4)$
\cite{BLT1} which can be seen as describing the $D=3$ quantum superconformal symmetry or
$D=4$ quantum anti-de-Sitter supersymmetry.

The quantum deformations of relativistic supersymmetries are described by Hopf-algebraic
deformations of the Poincar\'{e} superalgebra. Such quantum deformations are classified
by super-Poincar\'{e} Poisson structures which are given by classical $r$-matrices. In
the case of the Poincar\'{e} algebra complete classification of classical $r$-matrices,
which do satisfy the homogeneous classical Yang-Baxter equation, was obtained by S.
Zakrzewski in \cite{Z1}. In the case of  Poincar\'{e} superalgebra even partial
classification of corresponding $r$-matrices has not been achieved. In this paper we
undertake such a task. We shall extend supersymmetrically part of the Zakrzewski's list
by an addition to the Poincare classical $r$-matrices the terms containing supercharges.
Moreover these extended $r$-matrices can be presented as a sum of the so-called
subordinated $r$-matrices which are of super-Abelian and super-Jordanian type. This
subordination structure allows to construct a sequence of quantizations described by the
product of twists providing complete quantum deformation. The twists considered below are
super-extensions of the twists obtained recently in \cite{T1}.

\setcounter{equation}{0}
\section{Supertwists $-$ general considerations}

Let $r$ be a classical $r$-matrix of a Lie superalgebra $\mathfrak{g}$ ($r\in\,
\stackrel{2} \wedge\mathfrak{g}$) satisfying the classical Yang--Baxter equation (CYBE)
\begin{eqnarray}\label{p1}
[r^{12},\,r^{13}+\,r^{23}] + [r^{13},\,r^{23}]\!\!&=\!\!&\Omega~,
\end{eqnarray}
where $\Omega$ is $\mathfrak{g}$-invariant element, $\Omega\in(\stackrel{3}\wedge
\mathfrak{g})_{\mathfrak{g}}$. We consider two types of the classical $r$-matrices and
corresponding twists.

{\it a) Graded Abelian twist.} Let the classical $r$-matrix $r=r_{sA}^{}$ has the
form\footnote{We do not consider here situation when $r$-matrix contains terms of odd
degree, i.e. the terms of form $x_i\wedge u_k$, where $\deg{x_i}=0$, $\deg{u_k}=1$. In a
case of the such mixed terms we need to introduce odd (fermionic) parameters of
deformation in the $r$-matrix (for example see \cite{KS}; for general discussion see
\cite{T3}).}
\begin{eqnarray}\label{p2}
r_{sA}^{}\!\!&=\!\!&\sum_{i=1}^{n}y_{i}\wedge x_{i}+\sum_{i=1}^{m}v_{k}\wedge u_{k}~,
\end{eqnarray}
where all elements $x_i, y_i$ $(i=1,\ldots,n)$ are even (bosonic) and they commute among
themselves, all elements $u_k, v_k$ $(k=1,\ldots,m)$ are odd (fermionic) and they
anti-commute,  the elements $x_i, y_i$  do commute with the elements $u_k, v_k$. Moreover
the symbol ''$\wedge$'' for bosonic and fermionic elements is defined as follows
\begin{eqnarray}\label{p3}
y_{i}\wedge x_{i}\!\!&:=\!\!&y_{i}\otimes x_{i}-x_{i}\otimes y_{i}~,\quad\;\; {\rm
for}\;\; \deg{x_i}= \deg {y_i}=0~,
\\[7pt]\label{p4}
v_{k}\wedge u_{k}\!\!&:=\!\!&v_{k}\otimes u_{k}+u_{k}\otimes v_{k}~,\quad {\rm for}\;\;
\deg{u_k}= \deg {v_k}=1~.
\end{eqnarray}
The $r$-matrix (\ref{p2}) is called of super-Abelian type. The corresponding twist is
given as follows
\begin{eqnarray}\label{p5}
F_{r_{sA}^{}}\!\!&=\!\!&\exp\frac{\tilde{r}_{sA}^{}}{2}=
\exp\Bigl(\frac{1}{2}\sum_{i=1}^{n}x_{i}\wedge y_{i}+\frac{1}{2}\sum_{k=1}^{m}u_{k}\wedge
v_{k}\Bigr)~.
\end{eqnarray}
This twisting two-tensor $F:=F_{r_{A}^{}}$ satisfies the cocycle
equation \cite{D2}
\begin{equation}\label{p6}
F^{12}(\Delta\otimes{\rm id})(F)\;=\;F^{23}({\rm id}\otimes\Delta)(F)~,
\end{equation}
and the "unital" normalization condition
\begin{equation}\label{p7}
(\epsilon \otimes{\rm id})(F)\;=\;({\rm id}\otimes\epsilon )(F)\;=\;1~.
\end{equation}
The twisting element $F$ defines the deformation of universal enveloping algebra
$U(\mathfrak{g})$ considered as a Hopf algebra. The new deformed coproducts and antipodes
are given as follows
\begin{equation}\label{p8}
\Delta^{(F)}(a)\;=\;F\Delta(a)F^{-1}~,\qquad S^{(F)}(a)=uS(a)u^{-1}
\end{equation}
for any $a\in U(\mathfrak{g})$, where $\Delta(a)$ is a coproduct
before twisting, and
\begin{eqnarray}\label{p9}
u\!\!&=\!\!&\sum_i f^{(1)}_{i}S(f^{(2)}_i)
\end{eqnarray}
if $F=\sum_i f^{(1)}_i\otimes f^{(2)}_i$.

{\it b) Extended super Jordanian twist.} Let the classical $r$-matrix
$r=r_{J_{n|m}}^{}(\xi)$ has the following form\footnote{Here introduction of the
deformation parameter $\xi$ is a matter of convenience.}
\begin{equation}\label{p10}
r_{J_{n|m}}^{}(\xi)\;=\;\xi\,\Bigl(\sum_{\nu=0}^{n}y_{\nu}\wedge
x_{\nu}+\sum_{k=1}^{m}v_{k}\wedge u_{k}\Bigr)~,
\end{equation}
where the elements $x_\nu,y_\nu$ $(\nu=0,1,\ldots, n)$ and $u_k,v_k$
$(k=1,2,\ldots,m)$ satisfy the relations:
\begin{equation}\label{p11}
\begin{array}{rcl}
[x_{0},\,y_{0}]\!\!&=\!\!&y_{0}~,\qquad\quad
[x_{0},\,x_{i}]\,=\,t_{i}x_{i}~,\qquad\qquad\quad\;[x_{0},y_{i}]\,=\,(1-t_{i})y_{i}~,
\\[7pt]
[x_{i},\;y_{j}]\!\!&=\!\!&\delta_{ij}y_{0}~,\qquad[x_{i},\,y_{j}]\,=\,
[y_{i},y_{j}]\,=\,0~,\qquad[y_{0},x_{j}]\,=\,[y_{0},y_{j}]=0~,
\\[7pt]
[x_{0},u_{k}]\!\!&=\!\!&t_{k}'u_{k}~,\qquad[x_{0},\,v_{k}]\,=\,(1-t_{k}')v_{k}~,\qquad\;\,
[y_{0},u_{k}]\,=\,[y_{0},v_{k}]=0~,
\\[7pt]
[x_{i},u_{k}]\!\!&=\!\!&[x_{i},v_{k}]\;=\;[y_{i},u_{k}]\;\,=\;\,[y_{i},v_{k}]\;=\;0~,
\\[7pt]
\{u_{k},v_{l}\}\!\!\!&=\!\!\!&\delta_{kl}y_{0}~,\quad\;\;\{u_{k},u_{l}\}\;=\;
\{v_{k},v_{l}\}\;=\;0~,
\end{array}
\end{equation}
$(i,j=1,2\ldots,n)$, $(k,l=1,2\ldots,m)$, $(t_{i},t_{k}'\in{\mathbb C})$. Such an
$r$-matrix is called of super-Jordanian type and it is easy to verify that the two-tensor
(\ref{p10}) indeed satisfies the homogenous classical Yang-Baxter equation (\ref{p1})
(with $\Omega=0$), provided the elements $x_\nu,y_\nu$ $(\nu=0,1,\ldots,n)$ and $u_k,v_k$
$(k=1,2,\ldots,m)$ satisfy the relations (\ref{p11}). The corresponding twist is given as
follows \cite{T3,T4}
\begin{eqnarray}\label{p12}
F_{r_{J_{n|m}}^{}}\!\!&=\!\!&\exp\Big(\xi\sum\limits_{k=1}^{m}u_{k}\otimes
v_{k}\;e^{-2(1-t_{k}')\sigma}+ \xi\sum\limits_{i=1}^{n}x_{i}\otimes
y_{i}\;e^{-2(1-t_{i})\sigma}\Bigr)~, \exp(2x_{0}^{}\otimes\sigma),
\end{eqnarray}
where $\sigma:=\frac{1}{2}\ln(1+\xi y_{0})$.

{\it Remark.} The bosonic part $r_{J_{n}}^{}(\xi)=\xi\sum_{\nu=0}^{n}y_{\nu}\wedge
x_{\nu}$ of (\ref{p10}) is the classical $r$-matrix of Jordanian type and the
corresponding twist of Jordanian type is given by the formula (\ref{p12}) for $m=0$.

Let $r$ be an arbitrary $r$-matrix of the superalgebra $\mathfrak{g}$ and let
$\mathop{\rm Sup}(r)$ be a support of $r$~\footnote{The support $\mathop{\rm Sup}(r)$ is
a subalgebra of $\mathfrak{g}$ generated by the elements $\{x_i,y_i;u_k,v_k\}$ if
$r=\sum_{i}y_i\wedge x_i+\sum_{k}v_k\wedge u_k$.}. We recall the useful notion of
subordination  \cite{T1,T2}.
\begin{definition}
Let $r^{}_1$ and $r^{}_2$ be two arbitrary classical $r$-matrices. We say that $r^{}_2$
is subordinated to $r^{}_1$, $r^{}_1\succ r^{}_2$, if $\delta_{r^{}_1}(\mathop{\rm
Sup}(r^{}_2))=0$, i.e.
\begin{equation}\label{p13}
\delta_{r_{1}^{}}(x)\;:=\;[x\otimes1+1\otimes x,\,r_{1}^{}]\;=\;0~, \quad \forall x\in
\mathop{\rm Sup}(r_2)~.
\end{equation}
\end{definition}
If $r^{}_1\succ r^{}_2$ then $r=r^{}_1+r^{}_2$ is also a classical $r$-matrix. The
subordination enables us to construct a correct sequence of quantizations. For instance,
if the $r$-matrix of super-Jordanian type (\ref{p10}) is subordinated to the $r$-matrix
of super-Abelian type (\ref{p2}), $r_{\!sA}^{}\succ r_{\!J_{n|m}}^{}$, then the total
twist corresponding to the resulting $r$-matrix $r=r_{\!sA}^{}+r_{\!J_{n|m}}^{}$ is given
as follows
\begin{eqnarray}\label{p14}
F_{r}\!\!&=\!\!&F_{r_{\!J_{n|m}}^{}}F_{r_{\!sA}^{}}.
\end{eqnarray}

\setcounter{equation}{0}
\section{Quantum deformations of super-Poincar\'{e} algebra}

The $D=4$ Poincar\'{e} algebra ${\mathcal{P}}(3,1)$ is described by ten generators:\\
(i)  six-dimensinal Lorentz algebra
$\mathfrak{o}(3,1)$ with the generators  $M_i$, $N_i$ ($i=1,2,3$):
\begin{eqnarray}\label{sP1}
[M_i,\,M_j ]\ =\ \imath\,\epsilon_{ijk}\,M_k ~,\qquad [M_i,\,N_j]\
=\ \imath\,\epsilon_{ijk}\,N_k ~,\qquad [N_i,\,N_j]\ =\ -\imath\,
\epsilon_{ijk}\,M_k~
\end{eqnarray}
(ii) Abelian fourmomenta $P_0$, $P_j$ $(j=1,2,3)$ with the standard commutation
relations:
\begin{eqnarray}\label{sP2}
[M_j,\,P_k]\!\!&=\!\!&\imath\,\epsilon_{jkl}\,P_l~,\qquad
[M_j,\,P_0]\;=\;0~,
\\[5pt]\label{P3}
[N_j,\,P_k]\!\!&=\!\!&-\imath\,\delta_{jk}\,P_0~,\quad\;\;[N_j,\,P_0]\;=\;-\imath\,
P_j^{}~.
\end{eqnarray}
The physical generators of the Lorentz algebra $\mathfrak{o}(3,1)$, $M_i$, $N_i$ ($i =
1,2,3$), are related with the canonical basis $h,h',e_{\pm},e'_{\pm}$ as follows
\begin{eqnarray}\label{sP4}
h\!\!&=\!\!&\imath\, N_3~,\qquad e_{\pm}\;=\;\imath\, (N_1\pm\,M_2),
\\[5pt]\label{sP5}
h'\!\!&=\!\!&\imath\, M_3~,\qquad e'_{\pm}\;=\;\imath\, (M_1\mp
N_2)~,
\end{eqnarray}
where the generators (\ref{sP4}-\ref{sP5}) satisfy the following
nonvanishing commutation relations:
\begin{eqnarray}\label{sP6}
&[h,\,e_{\pm}^{}]\;=\;\pm e_{\pm}^{}\,,\qquad [e_{+}^{},\,e_{-}^{}]\;=\;2h~,
\\[5pt]\label{sP7}
&[h,\,e'_{\pm}]\;=\;\pm e'_{\pm}~,\quad\;\; [h',\,e_{\pm}]\;=\;\pm e'_{\pm}~,\quad\;\;
[e_{\pm}^{},\,e'_{\mp}]\;=\;\pm2h'~,
\\[5pt]\label{sP8}
& [h',\,e'_{\pm}]\;=\;\mp e_{\pm}^{}~,\qquad
[e'_{+},\,e'_{-}]\;=\;-2h~.
\end{eqnarray}
Moreover one can introduce the reality structure as follows
\begin{equation}\label{sP9}
a^*\;=\;-a\qquad (\forall\;a\;\in\;\mathfrak{o}(3,1))~.\textbf{}
\end{equation}
The subalgebra generated by the four momenta $P_0$, $P_j$ $(j=1,2,3)$ will be denoted by
$\mathbf{P}$ and we also set $P_{\pm}:=P_{0}\pm P_{3}$.

S.~Zakrzewski has shown in \cite{Z1} that each classical $r$-matrix,
$r\in\mathcal{P}(3,1) \wedge\mathcal{P}(3,1)$,  has the following decomposition
\begin{equation}\label{sP10}
r=a+b+c~,
\end{equation}
where $a\in\mathbf{P}\wedge\mathbf{P}$, $b\in\mathbf{P}\wedge\mathfrak{o}(3,1)$,
$c\in\mathfrak{o}(3,1)\wedge\mathfrak{o}(3,1)$ satisfy the following relations
\begin{eqnarray}\label{sP11}
[[c,c]]\!\!&=\!\!&0~,
\\[3pt]\label{sP12}
[[b,c]]\!\!&=\!\!&0~,
\\[3pt]\label{sP13}
2[[a,c]]+[[b,b]]\!\!&=\!\!&t\Omega\quad (t\in \mathbb{R},\;\Omega\neq0)~,
\\[3pt]\label{sP14}
[[a,b]]\!\!&=\!\!&0~
\end{eqnarray}
and $[[\cdot,\cdot]]$ means the Schouten bracket. Moreover a complete list of the
classical $D=4$ $r$-matrices for the cases $c\neq0$ and $c=0$, $t=0$ is
known.\footnote{Classification of the $r$-matrices for the case $c=0$, $t\neq0$ is still
not complete.} The results are presented in the following table taken from
\cite{Z1}:\\[10pt]
{
\begin{tabular}{ccccc}
\hline $c$ & $b$ & $a$ & $\#$ & $N$\\
\hline $\gamma h'\wedge h$ & $0$ & $\alpha P_{+}\wedge P_{-}+\tilde{\alpha}P_{1}\wedge
P_{2}$ & $2$ & $1$\\
\hline $\gamma e'_{+}\wedge e_{+}$ & $\beta_{1}b_{P_{+}}^{}+\beta_{2}P_{+}\wedge h'$ &
$0$ & $1$ & $2$\\
$$ & $\beta_{1} b_{P_{+}}^{}$ & $\alpha P_{+}\wedge P_{1}$ & $1$ & $3$\\
$$ & $\gamma\beta_{1}(P_{1}\wedge e_{+}+P_{2}\wedge e'_{+})$ & $P_{+}\wedge(\alpha_{1}P_{1}\!+
\alpha_{2}P_{2})-\gamma\beta_{1}^2P_{1}\wedge P_{2}$ & $2$ & $4$\\
\hline $\gamma(h\wedge e_{+}$ & $$ & $$ & $$ & $$\\
$-h'\wedge e'_{+})$ & $0$ & $0$ & $1$ & $5$\\
$+\gamma_{1}e'_{+}\wedge e_{+}$ & $$ & $$ & $$ & $$\\
\hline $\gamma h\wedge e_{+}$ & $\beta_{1}b_{P_{2}}^{}+\beta_{2}P_{2}\wedge e_{+}$ & $0$
& $1$ & $6$\\
\hline $0$ & $\beta_{1}b_{P_{+}}^{}+\beta_{2}P_{+}\wedge h'$ & $0$ & $1$ & $7$\\
$$ & $\beta_{1}b_{P_{+}}^{}+\beta_{2}P_{+}\wedge e_{+}$ & $0$ & $1$ & $8$\\
$$ & $P_{1}\wedge(\beta_{1}e_{+}+\beta_{2}e'_{+})\,+$ & $\alpha P_{+}\wedge P_{2}$ &
$2$ & $9$\\
$$ & $\beta_{1}P_{+}\wedge(h+\chi e_{+}),\;\chi=0,\pm1$ & $$ & $$ & $$\\
$$ & $\beta_{1}(P_{1}\wedge e'_{+}+P_{+}\wedge e_{+})$ & $\alpha_{1}P_{-}\wedge
P_{1}+\alpha_{2} P_{+}\wedge P_{2}$ & $2$ & $10$\\
$$ & $\beta_{1} P_{2}\wedge e_{+}$ & $\alpha_{1}P_{+}\wedge P_{1}+\alpha_{2}P_{-}\wedge P_{2}$ &
$1$ & $11$\\
$$ & $\beta_{1} P_{+}\wedge e_{+}$ & $P_{-}\!\wedge(\alpha P_{+}\!+\!\alpha_{1}P_{1}\!+\!\alpha_{2}P_{2})\!+
\tilde{\alpha} P_{+}\!\wedge P_{2}$ & $3$ & $12$\\
$$ & $\beta_{1} P_{0}\wedge h'$ & $\alpha_{1}P_{0}\wedge P_{3}+\alpha_{2}P_{1}\wedge P_{2}$ &
$2$ & $13$\\
$$ & $\beta_{1} P_{3}\wedge h'$ & $\alpha_{1}P_{0}\wedge P_{3}+\alpha_{2}P_{1}\wedge P_{2}$ &
$2$ & $14$\\
$$ & $\beta_{1} P_{+}\wedge h'$ & $\alpha_{1}P_{0}\wedge P_{3}+\alpha_{2}P_{1}\wedge P_{2}$ &
$1$ & $15$\\
$$ & $\beta_{1} P_{1}\wedge h$ & $\alpha_{1}P_{0}\wedge P_{3}+\alpha_{2}P_{1}\wedge P_{2}$ &
$2$ & $16$\\
$$ & $\beta_{1} P_{+}\wedge h$ & $\alpha P_{1}\wedge P_{2}+\alpha_{1}P_{+}\wedge P_{1}$ &
$1$ & $17$\\
$$ & $P_{+}\wedge(\beta_{1} h+\beta_{2} h')$ & $\alpha_{1} P_{1}\wedge P_{2}$ & $1$ & $18$\\
\cline{2-5}
$$ & $0$ & $\alpha_{1} P_{1}\wedge P_{+}$ & $0$ & $19$\\
$$ & $$ & $\alpha_{1} P_{1}\wedge P_{2}$ & $0$ & $20$\\
$$ & $$ & $\alpha_{1} P_{0}\wedge P_{3}+\alpha_{2}P_{1}\wedge P_{2}$ & $1$ & $21$\\
\hline
\end{tabular}
}
\begin{center}
{\bf Table 1.} Normal forms of $r$ for $c\neq0$ or $t=0$.
\end{center}
where $b_{P_{+}}^{}$ and $b_{P_{2}}^{}$ are given as follows:
\begin{eqnarray}\label{sP15}
b_{P_{+}}^{}\!\!&=\!\!&P_{1}\wedge e_{+}-P_{2}\wedge e'_{+}+P_{+}\wedge h~,
\\[5pt]\label{sP16}
b_{P_{2}}^{}\!\!&=\!\!&2P_{1}\wedge h'+P_{-}\wedge e'_{+}-P_{+}\wedge e'_{-}~.
\end{eqnarray}
The Table 1 contains 21 cases labelled by the number $N$ in the last column. In the forth
column (labelled by $\#$) there is indicated the number of essential parameters defining
given deformation. This number is smaller than the number of parameters occurring in the
Table 1. In particular, we introduced an additional parameter $\gamma$ in the component
$a$ (in the cases 2, 3, 4, 5, 6) the parameter $\beta_1^{}$ in the component $b$ (in the
cases 7--18) and the parameter $\alpha_1^{}$ in the component $a$ (in the cases
19--21)\footnote{In the paper by S. Zakrzewski \cite{Z1} all these additional parameters
are equal to 1 and the numbers in the forth column correspond to this situation.}. The
reduction of the number of parameters to the essential ones is achieved by taking into
account all possible automorphisms of the Poincar\'{e} algebra ${\mathcal{P}}(3,1)$ (see
details in \cite{Z1}).

The super-Poincar\'{e} algebra ${\mathcal P}(3,1|1)$ is generated by the algebra
$\mathcal{P}(3,1)$ and four real supercharges $Q_\alpha^{}$ $(\alpha=\pm1,\pm2)$) with
the anticommutation relations
\begin{equation}
\begin{array}{rcl}\label{sP18}
\{Q^{\pm}_{\alpha},\,Q^{\pm}_{\beta}\}\!\!&=\!\!&0~,\qquad
\{Q^{+}_{\alpha},\,Q^{-}_{\beta}\}=2\bigl( \delta_{\alpha\beta}^{}\,P_0^{}+
(\sigma_j^{})_{\alpha\beta}^{}\,P_j^{}\bigr)~,
\end{array}
\end{equation}
where $\sigma_{\!j}^{}$ $(j=1,2,3)$ are the standard $2\times2$ $\sigma$-matrices, and
moreover
\begin{equation}
\begin{array}{rcl}\label{sP17}
[M_j^{},Q^{\pm}_{\alpha}]\!\!&=\!\!&\displaystyle\pm\frac{1}{2}(\sigma_j^{\pm})_
{\alpha\beta}^{}\,Q^{\pm}_{\beta},
\\[9pt]
[N_j^{},Q^{\pm}_{\alpha}]\!\!&=\!\!&\displaystyle
-\frac{\imath}{2}(\sigma_j^{\pm})_
{\alpha\beta}^{}\,Q^{\pm}_{\beta},\qquad
[P_\mu^{},Q^{\pm}_{\alpha}]\,=\,0~.
\end{array}
\end{equation}
Here $\sigma_j^{+}\equiv\sigma_j^{}$ and $\sigma_j^{-}$ denotes the complex conjugate of
$\sigma_j^{}$~.

It is clear that all classical $r$-matrices satisfying the homogeneous YBE are
$r$-matrices for the super-Poincar\'{e} algebra and the twists constructed in
\cite{T3,T4} can be used for the derivation of explicit super-Poincare twists. It turns
out that it is possible to extend supersymmetrically the Zakrzewski's classification for
the Poincare superalgebra by an addition of terms expressed in terms of supercharges.
Moreover these extended $r$-matrices can be presented as a sum of subordinated
$r$-matrices which are of super-Abelian and super-Jordanian types. We consider below
several examples of such superextensions and give the corresponding twists describing the
quantum deformations for the first six $r$-matrices from the Table 1, with $c\neq 0$,
$t=0$. These twist quantizations are described by the  superextensions of the twists
obtained in \cite{T1}.

1). The $r$-matrix $r_1$ describing the superextension of $N=1$ in
{\bf Table 1}
\begin{eqnarray}\label{sP19}
r_{1}^{}\!\!&=\!\!&\gamma h'\wedge h+\alpha P_{+}\wedge P_{-}+\tilde{\alpha} P_{1}\wedge
P_{2}
\end{eqnarray}
has the following superextension:
\begin{eqnarray}\label{sP20}
r_{1s}^{}\!\!&=\!\!&\gamma\,h'\wedge h+\alpha\,P_{+}\wedge P_{-}+\tilde{\alpha}\,
P_{1}\wedge P_{2}+\eta\,Q^{+}_{2}\wedge Q^{+}_{1}~.
\end{eqnarray}
The $r$-matrix (\ref{sP19}) can be presented as a sum of two
subordinated $r$-matrices of super-Abelian and Abelian types
\begin{eqnarray}\label{sP21}
\begin{array}{rcl}
r_{1s}\!\!&=\!\!&r_{1s}'+r_{1}''~,\quad r_{1s}'\,\succ\,r_{1}''~,
\\[7pt]
r_{1s}'\!\!&=\!\!&\alpha\,P_{+}\wedge P_{-}+\tilde{\alpha}\,P_{1}\wedge
P_{2}+\eta\,Q^{+}_{2}\wedge Q^{+}_{1}~,
\\[7pt]
r_{1}''\!\!&=\!\!&\gamma\,h'\wedge h~.
\end{array}
\end{eqnarray}
The complete twist defining quantization
is the ordered product of super-Abelian and
Abelian twists
\begin{eqnarray}\label{sP22}
F_{r_{1s}^{}}\!\!\!&=\!\!\!&F_{r_{1}''}\,F_{r_{1s}'}\,=\,\exp\bigl(\gamma\,h\wedge
h'\bigr) \exp\bigl(\alpha P_{-}\wedge P_{+}+\tilde{\alpha}\,P_{2}\wedge
P_{1}+\eta\,Q^{+}_{1}\wedge Q^{+}_{2}\bigr).
\end{eqnarray}

2). The $r$-matrix $r_2$, describing the superextension of $N=2$ in
{\bf Table 1}
\begin{eqnarray}\label{sP23}
r_{2}^{}\!\!&=\!\!&\gamma\,e'_{+}\wedge e_{+}+\beta_{1}\,(P_{1}\wedge e_{+}-P_{2}\wedge
e'_{+}+P_{+}\wedge h)+\beta_{2}\,P_{+}\wedge h'~,
\end{eqnarray}
admits the superextension
\begin{eqnarray}\label{sP24}
\begin{array}{rcl}
r_{2s}^{}\!\!&=\!\!&\gamma\,\displaystyle e'_{+}\wedge e_{+}+\beta_{1}(P_{1}\wedge
e_{+}+P_{2}\wedge e'_{+}+P_{+}\wedge h)+\beta_{2}\,P_{+}\wedge
h'+\frac{\beta_{1}}{2}\,Q_{1}^{-}\wedge Q_{1}^{+},
\end{array}
\end{eqnarray}
The $r$-matrix $r_{2s}^{}$ can be presented as a sum of three subordinated $r$-matrices
where one of them is of super-Jordanian type and two are of Abelian type
\begin{eqnarray}\label{sP25}
\begin{array}{rcl}
r_{2s}^{}\!\!\!&=\!\!\!&r_{2s}'+r_{2}''+r_{2}''',\quad r_{2s}'\;\succ\, r_{2}''~,\quad
r_{2s}'+r_{2}''\;\succ\,r_{2}'''~,
\\[9pt]
r_{2s}'\!\!&=\!\!&\displaystyle\beta_{1}\,\bigl(\frac{1}{2}\,Q_{1}^{-}\wedge
Q_{1}^{+}+P_{1}\wedge e_{+}-P_{2}\wedge e'_{+}+P_{+}\wedge h\bigr)~,
\\[12pt]
r_{2}''\!\!&=\!\!&\gamma\,e'_{+}\wedge e_{+}~,\qquad r_{2}'''\;=\;\beta_{2}\,P_{+}\wedge
h'~.
\end{array}
\end{eqnarray}
The corresponding twisting two-tensor is given by the following
formula
\begin{eqnarray}\label{sP26}
F_{r_{2s}^{}}\!\!&=\!\!&F_{r_{2s}'''}\,F_{r_{2}''}\,F_{r_{2}'}~,
\end{eqnarray}
where
\begin{eqnarray}\label{sP27}
\begin{array}{rcl}
F_{r_{2s}^{'}}\!\!&=\!\!&\displaystyle\exp\Bigr(\beta_{1}\,\bigl(\frac{1}{2}\,Q_{1}^{+}
\otimes Q_{1}^{-}e^{-\frac{1}{2}\sigma_{+}}+e_{+}^{}\otimes P_{1}-e_{+}'\otimes
P_{2}\bigr)\Bigr)\exp(2h\otimes\sigma_{+})~,
\\[14pt]
F_{r_{2}^{''}}\!\!&=\!\!&\exp(\gamma\,e_{+}\wedge e'_{+})~,
\\[7pt]
F_{r_{2}'''}\!\!&=\!\!&\displaystyle\exp\Bigl(\frac{\beta_2}{\beta_1}\,h'\wedge
\sigma_{+}\Bigr)~.
\end{array}
\end{eqnarray}
Here and below we set $\sigma_+:=\frac{1}{2}\ln(1+\beta_{1}P_{+})$.

It should be noted that the first formula (\ref{sP27}) containing the supercharges
$Q_{\alpha}^{\pm}$ is related with the result presented in \cite{BLT1}

3). The $r$-matrix $r_3$, describing the superextension of $N=3$ in
{\bf Table 1}.
\begin{eqnarray}\label{sP28}
\begin{array}{rcl}
r_3\!\!&=\!\!&\gamma\,e'_{+}\wedge e_{+}+\beta_{1}\,(P_{1}\wedge e_{+}-P_{2}\wedge e'_{+}
+P_{+}\wedge h)+\alpha\,P_{+}\wedge P_{1}~,
\end{array}
\end{eqnarray}
also admits the superextension in analogy to the previous case
\begin{eqnarray}\label{sP29}
\begin{array}{rcl}
r_{3s}^{}\!\!&=\!\!&\displaystyle\gamma\,e'_{+}\wedge e_{+}+\beta_{1}\,\bigr(P_{1}\wedge
e_{+}-P_{2}\wedge e'_{+}+P_{+}\wedge h\bigl)+\alpha\,P_{+}\wedge
P_{1}+\frac{\beta_{1}}{2}\,Q_{1}^{-}\wedge Q_{1}^{+},
\end{array}
\end{eqnarray}
This $r$-matrix can be presented as a sum of three subordinated $r$-matrices where one of
them is of super-Jordanian type and two are of Abelian type
\begin{eqnarray}\label{sP30}
\begin{array}{rcl}
r_{3s}^{}\!\!&=\!\!&r_{3s}'+r_{3}''+r_{3}'''~,\quad r_{3s}'\;\succ\, r_{3}''~,\quad
r_{3s}'+r_{3}''\;\succ\,r_{3}'''~,
\\[11pt]
r_{3s}'\!\!&=\!\!&\displaystyle\beta_{1}\Bigl(\frac{1}{2}\,Q_{1}^{-}\wedge
Q_{1}^{+}+P_{1}\wedge\bigl(e_{+}-\frac{\alpha}{\beta_{1}}\,P_{+}\bigr)-P_{2}\wedge
e'_{+}+P_{+}\wedge h\Bigr)~,
\\[11pt]
r_{3}''\!\!&=\!\!&\displaystyle\gamma e'_{+}\wedge\bigl(e_{+}-
\frac{\alpha}{\beta_{1}}P_{+}\bigr)~,
\\[11pt]
r_{3}'''\!\!&=\!\!&\displaystyle\frac{\gamma\alpha}{\beta_{1}}e'_{+}\wedge P_{+}~.
\end{array}
\end{eqnarray}
The corresponding twist is given by the following formula
\begin{eqnarray}\label{sP31}
F_{r_3^{}}\!\!&=\!\!&F_{r_{3}'''}\,F_{r_{3}''}\,F_{r_{3s}'}~,
\end{eqnarray}
where
\begin{eqnarray}\label{sP32}
\begin{array}{rcl}
F_{r_{3s}'}\!\!&=\!\!&\displaystyle\exp\Bigl(\beta_{1}\Bigl(\frac{1}{2}\,Q_{1}^{+}\otimes
Q_{1}^{-}e^{-\frac{1}{2}\sigma_{+}}+e_{+}^{}- \frac{\alpha}{\beta_{1}}\,P_{+})\otimes
P_{1}-e'_{+}\otimes P_{2}\Bigr)\Bigr) \exp(2h\otimes\sigma_{+})~,
\\[12pt]
F_{r_{3}''}\!\!&=\!\!&\displaystyle\exp\Bigl(\gamma\,\bigl(e_{+}-
\frac{\alpha}{\beta_{1}}\,P_{+}\bigr)\wedge e'_{+}\Bigr)~,
\\[12pt]
F_{r_{3}''}\!\!&=\!\!&\displaystyle\exp\Bigl(
\frac{\gamma\alpha}{\beta_{1}^2}\,\sigma_{+}\wedge e'_{+}\Bigr)~.
\end{array}
\end{eqnarray}

4). The $r$-matrix $r_4$, describing the superextension of $N=4$ in
{\bf Table 1}.
\begin{eqnarray}\label{sP33}
\begin{array}{rcl}
r_{4}^{}\!\!&=\!\!&\gamma(e'_{+}\wedge e_{+}+\beta_{1}P_{1}\wedge e_{+}+\beta_{1}
P_{2}\wedge e'_{+}\!-\beta_{1}^{2}P_{1}\wedge P_{2})+P_{+}\wedge(\alpha_{1}P_{1}+
\alpha_{2}P_{2}),
\end{array}
\end{eqnarray}
has the following superextension
\begin{eqnarray}\label{sP34}
\begin{array}{rcl}
r_{4}^{}\!\!&=\!\!&\gamma\,(e'_{+}\wedge e_{+}+\beta_{1}P_{1}\,\wedge e_{+}+\beta_{1}\,
P_{2}\wedge e'_{+}\!-\beta_{1}^{2}P_{1}\wedge P_{2})\,+
\\[7pt]
&&+\,P_{+}\wedge(\alpha_{1}\,P_{1}+\alpha_{2}\,P_{2})+\eta\,Q^{+}_1\wedge Q^{+}_1,
\end{array}
\end{eqnarray}
This $r$-matrix can be written as a sum of two subordinated
$r$-matrices of super-Abelian and Abelian types
\begin{eqnarray}\label{sP35}
\begin{array}{rcl}
r_{4s}^{}\!\!&=\!\!&r_{4s}'+r_{4}''~,\quad r_{4s}'\;\succ\, r_{4}''~,
\\[9pt]
r_{4s}'\!\!&=\!\!&P_{+}\wedge(\alpha_{1}P_{1}+\alpha_{2}P_{2})+\eta\,Q^{+}_{1}\wedge
Q^{+}_{1}~,
\\[9pt]
r_{4}''\!\!&=\!\!&\gamma(e'_{+}+\beta_{1} P_{1})\wedge(e_{+}^{}-\beta_{1} P_{2})~.
\end{array}
\end{eqnarray}
The corresponding twist is given by the following formula
\begin{eqnarray}\label{sP36}
F_{r_{4s}^{}}\!\!&=\!\!&F_{r_{4}''}\,F_{r_{4s}'}~,
\end{eqnarray}
where
\begin{eqnarray}\label{sP37}
\begin{array}{rcl}
F_{r_{4s}^{'}}\!\!&=\!\!& \exp\bigr((\alpha_{1}P_{1}+\alpha_{2}P_{2})\wedge
P_{+}^{}+\eta\,Q^{+}_{1}\wedge Q^{+}_{1}\bigr)~,
\\[12pt]
F_{r_{4}^{''}}\!\!&=\!\!& \exp\bigr(\gamma(e_{+}-\beta_{1}P_{1})\wedge(e'_{+}+\beta_{1}
P_{2})\bigr)~.
\end{array}
\end{eqnarray}

It should be noted that the parameter $\beta_{1}$ can be removed by
the rescaling automorphism $\beta_{1}P_{\nu}\rightarrow P_{\nu}$
($\nu=0,1,2,3$).

5). The fifth $r$-matrix $r_5$ ($N=5$  in {\bf Table 1}) is the
$r$-matrix for the Lorentz algebra $\mathfrak{o}(3,1)$  which does
not admit any superextension. The explicit formulae for the
corresponding twist and algebra coproduts are calculated in
\cite{BLT2}.

6). The sixth  $r$-matrix $r_6$, describing the superextension of
$N=6$ in {\bf Table 1}
\begin{eqnarray}\label{sP38}
\begin{array}{rcl}
r_{6}^{}\!\!&=\!\!&\gamma\,h\wedge e_{+}+\beta_{1}\,(2P_{1}\wedge h'+P_{-}\wedge
e'_{+}-P_{+}\wedge e'_{-})+\beta_{2}\,P_{2}\wedge e_{+}~,
\end{array}
\end{eqnarray}
has the following superextension
\begin{eqnarray}\label{sP39}
\begin{array}{rcl}
r_{6s}^{}\!\!&=\!\!&\gamma\,h\wedge e_{+}+\beta_{1}\,(2P_{1}\wedge h'+P_{-}\wedge
e'_{+}-P_{+}\wedge e'_{-})+
\\[9pt]
&&+\displaystyle\,\beta_{2}\,P_{2}\wedge
e_{+}+\frac{\imath\beta_{1}}{4}\,(Q^{+}_{1}+Q^{-}_{1}) \wedge(Q^{+}_{2}-Q^{-}_{2})~,
\end{array}
\end{eqnarray}
This $r$-matrix can be presented as a sum of three subordinated $r$-matrices
\begin{eqnarray}\label{sP40}
\begin{array}{rcl}
r_{6s}\!\!&=\!\!&r_{s6}'+r_{6}''+r_{6}''',\quad r_{s6}'\,\succ\,r_{6}''~,\quad
r_{6}'+r_{6}''\,\succ\,r_{6}'''~,
\\[10pt]
r_{6s}'\!\!&=\!\!&\displaystyle\beta_{1}\Bigl(2P_{1}\wedge h'+P_{-}\wedge
e'_{+}-P_{+}\wedge e'_{-}+ \frac{\imath}{4}(Q^{+}_{1}+Q^{-}_{1})
\wedge(Q^{+}_{2}-Q^{-}_{2})\Bigr)~,
\\[10pt]
r_{6}''\!\!&=\!\!&\gamma h\wedge e_{+}~,
\\[10pt]
r_{6}'''\!\!&=\!\!&\beta_{2}P_{2}\wedge e_{+}~.
\end{array}
\end{eqnarray}
The $r$-matrix $r_{6}''$ is of Jordanian type, the $r$-matrix $r_{6}'''$ is of Abelian
type but the first $r$-matrix $r_{6s}'$ is neither of super-Abelian nor of
super-Jordanian type. One can check that $r$-matrix (\ref{sP39}) satisfies the
non-homogeneous classical Yang-Baxter equation (\ref{sP13}) with $t\neq0$. In terms of
the generators  $M_i$, $N_i$ ($i=1,2,3$) the $r$-matrix $r_{6s}'$ has the form
\begin{eqnarray}\label{sP41}
r'_{6s}\!\!&=\!\!&2\imath\beta_{1}\Bigl(P_{1}\wedge M_{3}-P_{3}\wedge M_{1}-P_{0}\wedge
N_{2}+\frac{1}{8}(Q^{+}_{1}+Q^{-}_{1}) \wedge(Q^{+}_{2}-Q^{-}_{2})\Bigr)~.
\end{eqnarray}
Unfortunately the quantum deformation corresponding to the $r$-matrix (\ref{sP39}) can
not be described by a supertwist. It is quite plausible that the quantization of
(\ref{sP41}) can be obtained by particular contraction procedure from the $q$-deformation
$U_q(\mathfrak{osp}((1|4))$ of $D=4$ AdS superalgebra in analogy with the
$\kappa$-deformation of super-Poincar\'{e} algebras derived in \cite{LNS}.

We conclude that we presented superextensions for all Zakrzewski's classical $r$-matrices
when $c\neq0$ (see (\ref{sP10})) and we constructed the corresponding twists for the
cases $t=0$. Analogous results can be obtained for  classical $r$-matrices $r_{N}$
($7\leq N\leq 18$) in {\bf Table 1} with $c=0$, $t=0$.

\setcounter{equation}{0}
\section{Outlook}

In our paper we provided a supersymmetrization of the part of Zakrzewski's list of
Poincare algebra deformations, presented in the form of $r$-matrices (\cite{Z1}; see {\bf
Table 1}). It should be recalled that known examples of $D=4$ quantum deformations of
supersymmetries, e.g. standard $\kappa$-deformation \cite{LNS}
does not belong to the considered class, because they correspond in formula (\ref{sP10})
to the case $c=0$, $b\neq 0$. We see therefore as important physically task the
completion of the classification of quantum deformations of relativistic supersymmetries,
corresponding to cases with $c=0$ (see $7\leq N\leq 21$ in {\bf Table 1}). In particular,
one can consider the supersymmetrization of twisted versions of $\kappa$-deformed
Poincare algebra \cite{LL}.

In this presentation we consider only the Hopf-algebraic framework of quantum-deformed
$D=4$ Poincare supersymmetries and did not study its applications. It should be recalled
that in simplest case of Abelian twist there were already considered examples of deformed
field-theoretic models (see e.g. \cite{ABDMSW,ADMSW}). It should be pointed out that most
of the deformations of supersymmetric field theories were made without any reference to
twisting and Hopf algebra structure (see e.g. \cite{FL,Seib,FLM}). The next step in our
investigations will be as well  the results presented in Sec.3 in terms of physical
super-Poincare generators, to consider their superspace realizations and finally to
construct new dynamical models with quantum supersymmetries.

\newpage
\section*{Acknowledgments}
The paper has been supported by MNiSW grant NN202 318534 (A.B., J.L., V.N.T) and the
grant RFBR-08-01-00392 (V.N.T.), and the French National Research Agency grant
NT05-241455GIPM (V.N.T.).


\end{document}